\documentclass[letterpaper, 10 pt, conference]{ieeeconf}  % Comment this line out if you need a4paper

\IEEEoverridecommandlockouts                              % This command is only needed if
\usepackage{spconf,amsmath,graphicx}
\usepackage{makecell}
\usepackage{amsfonts}
\usepackage{xcolor}

\title{ \bf CAE-Transformer: transformer-based model to predict invasiveness of lung adenocarcinoma subsolid nodules from non-thin section 3D CT scans}
\name{\makecell{Shahin Heidarian$^{1}$,
Parnian Afshar$^{2}$,
Anastasia Oikonomou$^{3}$,\\
Konstantinos N. Plataniotis$^{4}$,
and Arash Mohammadi$^{2}$\thanks{This work was partially supported by the Natural Sciences and Engineering Research Council (NSERC) of Canada through the Create Grant RGPIN- 2016-04988.}}}
\address{$~^{1}$Department of Electrical and Computer Engineering, Concordia University, Montreal, Canada \\
$~^{2}$Concordia Institute for Information Systems Engineering, Concordia University, Montreal, Canada\\
$~^{3}$Department of Medical Imaging, Sunnybrook Health Sciences Centre, Toronto, Canada\\
$~^{4}$Department of Electrical and Computer Engineering, University of Toronto, Toronto, Canada}
\begin{document}
%\ninept
%
\maketitle
\thispagestyle{empty}
\pagestyle{empty}
%
%============================================================
\begin{abstract}
Lung cancer is the leading cause of mortality from cancer worldwide and has various histologic types, among which Lung Adenocarcinoma (LUAC) has recently been the most prevalent one.  The current approach to determine the invasiveness of LUACs is surgical resection, which is not a viable solution to fight lung cancer in a timely fashion. An alternative approach is to analyze chest Computed Tomography (CT) scans. The radiologists' analysis based on CT images, however, is subjective and might result in a low accuracy. In this paper, a transformer-based framework, referred to as the ``CAE-Transformer", is developed to efficiently classify LUACs using whole CT images instead of finely annotated nodules. The proposed CAE-Transformer can achieve high accuracy over a small dataset and requires minor supervision from radiologists. The CAE-Transformer utilizes an encoder to automatically extract informative features from CT slices, which are then fed to a modified transformer to capture global inter-slice relations and provide classification labels.  Experimental results on our in-house dataset of 114 pathologically proven Sub-Solid Nodules (SSNs) demonstrate the superiority of the CAE-Transformer over its counterparts, achieving an accuracy of 87.73\%, sensitivity of 88.67\%, specificity of 86.33\%, and AUC of 0.913, using a 10-fold cross-validation.
\vspace{-0.1in}
\newline

\indent \textit{Clinical relevance}—The proposed framework provides timely and accurate information about the invasiveness of lung cancer with minor supervision, which can lead to a proper treatment plan and reduce the risk of unnecessary or late surgeries.
\end{abstract}
%============================================================
%
%============================================================
%\begin{keywords}
%Lung Adenocarcinoma, Lung Nodule Invasiveness, Transformer, Subsolid Nodule, Self-Attention.
%\end{keywords}
%============================================================
%
%%%%%%%%%%%%%%%%%%%%%%%%%%%%%%%%%%%%%%%%%%%%%%
\vspace{-0.1in}
\section{Introduction} \label{sec:intro}
%%%%%%%%%%%%%%%%%%%%%%%%%%%%%%%%%%%%%%%%%%%%%%
Lung Cancer (LC) is the deadliest and least funded cancer worldwide~\cite{Kamath2019,Bray2018}. Non-small-cell LC is the major type of LC, and Lung Adenocarcinoma (LUAC) is the most prevalent histologic sub-type~\cite{Herbst2018}. Lung nodules manifesting as Ground Glass (GG) or Subsolid Nodules (SSNs) on CT have a higher risk of malignancy than other incidentally detected small solid nodules.  SNNs are often diagnosed as LUACs which are categorized according to their histology into three categories: pre-invasive lesions including atypical adenomatous hyperplasia (AAH) and adenocarcinoma in situ (AIS), minimally invasive (MIA), and invasive pulmonary adenocarcinoma (IPA)~\cite{Lai2021}. A timely and accurate attempt to differentiate the LUACs is of utmost importance to guide a proper treatment plan, as in some cases, a pre-invasive or minimally invasive SSN can be monitored with regular follow up CTs, whereas invasive lesions should undergo immediate surgical resection if they are deemed eligible. Most often, the SSN's types are diagnosed based on their pathological findings performed after surgical resection which is not desired for prior treatment planning. Currently, radiologists use chest Computed Tomography (CT) scans to assess the invasiveness of the SSNs based on their imaging findings and patterns.
%prior to making decisions regarding the appropriate treatment.
Such visual approaches, however, are time-consuming, subjective, and error-prone.

In this regard, many studies~\cite{Cui2020, Shao2020} have used high-resolution and thin-slice ($<1.5mm$) CT images (slices).  In practice, however,  lung nodules are mostly identified from CT scans performed for various clinical purposes,  acquired using routine standard or low dose scanning protocols with non-thin slice thicknesses (up to $5mm$)~\cite{Oikonomou2019}.
%which require longer analysis times, as well as more storage capacity and reconstruction time~\cite{Cui2020, Shao2020}.
Moreover,  recent lung cancer screening recommendations suggest using Low Dose CT scans (LDCT) with thicker slice-thicknesses (up to $2.5mm$)~\cite{Kazerooni2014}.
Capitalizing on the above discussion, the necessity of developing an automated classification framework that performs well regardless of technical settings has recently arisen among the research community.
\newline
\noindent
\textbf{Related Works:} In general, existing publications on the SSN classification and invasiveness assessment can be categorized into two main groups: (1) Radiomics-based and (2) Deep Learning-based frameworks~\cite{Gu2021}. In the former, a set of histogram-based, morphological, and clinical features are extracted from the CT images which are then analyzed using statistical or machine learning techniques such as the study conducted in Reference~\cite{Uthoff2019}.
As another example of such frameworks,  in Reference~\cite{Oikonomou2019}, a set of radiomics features are extracted from manually annotated nodules and used along with additional features obtained via the Functional Principal Component Analysis (FPCA) to train a linear logistic regression, achieving the accuracy of $81.0\%$ on a dataset of primary LUACs from non-thin CT scans of $109$ pathologically labeled SSNs.
%As another example of such frameworks, a histogram-based model is developed in Ref.~\cite{Oikonomou2019} to predict the invasiveness of primary adenocarcinoma SSNs from non-thin CT scans of $109$ pathologically labeled SSNs. In this study, a set of histogram-based and morphological features along with additional features extracted via the Functional Principal Component Analysis (FPCA) is fed to a linear logistic regression, achieving the accuracy of $81.0\%$ and Area Under the ROC Curve (AUC) of $0.91$.
Deep learning-based frameworks, on the other hand, extract informative features in an automated fashion. Existing deep models working with volumetric CT scans can be classified into two main categories:~(\textit{i}) The first group includes the 3D models (e.g., 3D CNN), which are supplied by the whole volume of images (i.e., all 2D slices) or a stack of all nodule patches (cropped images including nodules)~\cite{Liu2017}. Processing a large 3D dataset at once, however, demands more complex models, more computational resources, and larger labeled datasets, and; (\textit{ii}) The second approach, on the other hand, analyzes individual 2D CT slices or Regions of Interest (ROIs) in the first step, and utilizes an aggregation mechanism to represent characteristics of the whole volume~\cite{Heidarian2021,Heidarian2021a}. Due to the nature of 3D CT scans, which are essentially sequences of 2D images,  sequential deep models can be adopted to analyze them.
%there has recently been a surge of interest in the application of sequential models in diagnostic/prognostic tasks based on volumetric CT scans.
Conventional sequential models such as LSTM and RNN, however,  are incapable of capturing global context and dependencies between instances in sequential data and require huge computing resources. Transformer architecture~\cite{Vaswani2017}, on the other hand,  is a recently-proposed alternative sequential model based on the novel self-attention mechanism, which is capable of capturing global relations between instances while requiring far less computational resources compared to conventional LSTM and RNN networks. Transformers are also superior to their counterparts in terms of parallelization and dynamic attention.
\newline
\noindent
\textbf{Challenges and Contributions:}
Existing transformer-based models applicable in the image processing tasks such as Vision Transformer (ViT)~\cite{Dosovitskiy2020} and Convolutional Vision Transformer (CvT)~\cite{Wu2021} apply the self-attention function to the small patches in single 2D images and find the relation between them.  Analyzing a series of CT slices, however, requires a framework capable of capturing inter-slice relations.
In this study, we have developed an automated predictive framework based on the novel self-attention mechanism and the transformer encoder, referred to as the ``CAE-Transformer".
Unlike ViT and CvT, our proposed framework uses a Convolutional Auto-Encoder (CAE)~\cite{Masci2011} to extract informative features from CT slices in an unsupervised fashion and stack them to form a sequential feature map. The CAE is first pre-trained on the public LIDC-IDRI dataset, then fine-tuned on our in-house dataset. The obtained sequential feature maps are then used to provide the final predictions. As previously mentioned,  in the case of a volumetric CT scan,  beside  2D patterns, capturing inter-slice relations in the axial direction is of utmost importance, which is addressed in our proposed framework. To the best of our knowledge, this manuscript is the first one targeting the lung cancer invasiveness prediction task using a transformer-based model. It is also worth mentioning that most studies on lung cancer classification are based on the public LIDC-IDRI dataset~\cite{Armato2011}, in which nodule patches are manually annotated and used to train the models, which is a challenging and time-consuming task even for expert radiologists.  In this study,  however, we used a relatively small dataset without using the nodule annotations. In fact,  the need for exact tumor annotation is completely eliminated, and the model is supplied by the whole images with the evidence of tumor which are much easier to identify.
Another challenge is that the transformer architecture requires large training datasets to perform well,  which are hard to obtain in the medical domain.  As such,  particular modifications have been made to the transformer encoder's architecture as well as the pre-training and fine-tuning steps, making the model suitable for the small dataset. More specifically, the class token is removed, and the commonly used Global Average Pooling (GAP) layer at the top of the network has been replaced by a Global Max Pooling (GMP) layer.  Furthermore,  different training and pre-training approaches have been used in this study.  In particular,  label smoothing~\cite{Szegedy_2016_CVPR} is used during the training step, and only the auto-encoder part of the framework is pre-trained, not the transformer itself.  Besides,  only a few middle layers in the encoder-decoder network are pre-trained instead of the entire network,  and CT images are used for this purpose,  in contrast to other models which utilize natural images from the ImageNet dataset~\cite{Deng2009}.
%============================================================
%\setlength{\textfloatsep}{0pt}
\begin{figure}[h]
	\vspace{-0.15in}
    \centering
    \includegraphics[width=0.85\columnwidth]{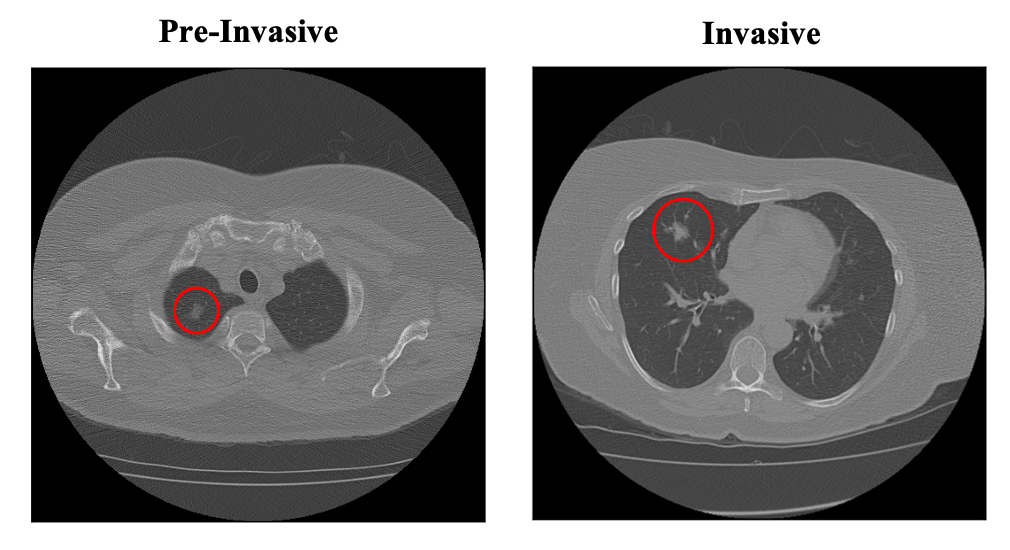}
    \caption{Sample pre-invasive and invasive adenocarcinomas.}
    \label{fig:sample-ct}
    \vspace{-0.2in}
\end{figure}
%============================================================
\section{CAE-Transformer Framework}
%============================================================
\subsection{Dataset}
In this study, we have used the dataset initially introduced in Reference~\cite{Oikonomou2019} and added five additional cases acquired from the same institution to further balance the dataset. This dataset contains volumetric CT scans of $114$ pathologically proven SSNs, identified and reviewed by $2$ experienced thoracic radiologists.
All SSN labels are provided after surgical resections.
SSNs are initially classified into three categories of pre-invasive lesions,  minimally invasive, and invasive pulmonary
adenocarcinoma. Following the original study~\cite{Oikonomou2019}, we have grouped the first two categories to represent the benign class with $58$ case and kept the invasive nodules as the malignant class, including $56$ cases.
In addition to the nodule labels, the CT slices with the evidence of a nodule are also specified by the radiologists.  Fig.~\ref{fig:sample-ct} shows two sample LUACs from this dataset.
%============================================================
\subsection{Lung Segmentation}
%============================================================
As the pre-processing step, we have utilized a well-trained U-Net-based lung segmentation model, introduced in Reference~\cite{Hofmanninger2020}, to extract the lung parenchyma from the CT scans. This approach has demonstrated a remarkable capability in enhancing the performance of models in previous studies~\cite{Heidarian2021,Mohammadi2021} by removing distracting components from the CT images. The extracted lung areas are then resized from $(512,512)$ to $(256,256)$ to reduce the complexity and memory allocation without significant loss of information.

%============================================================
\subsection{Convolutional Auto-Encoder (CAE)}
%============================================================
In order to represent CT images with compressed and informative feature maps, to be used as the input of the subsequent modules, we initially pre-trained a CAE on the public LIDC-IDRI dataset,  which contains $244,527$ CT images with or without the evidence of a nodule.
The designed CAE model consists of an encoder and a decoder part. The encoder is responsible for generating a compressed representation of the input image through a stack of 5 convolution and 5 max-pooling layers followed by a fully-connected layer with the size of $256$, while the decoder part attempts to reconstruct the original image using the feature representation generated by the encoder. By minimizing the MSE error between the original and the reconstructed image, the CAE learns to produce highly informative feature representations for the input images. Finally, the pre-trained model is fine-tuned on our in-house dataset.

%============================================================
\subsection{Multi-Head Self-Attention Mechanism}
%============================================================
The transformer model is the building block of the CAE-Transformer framework and uses a novel self-attention mechanism to capture global dependencies among instances in the input sequence with a high parallelization capability.
The self-attention mechanism is based on a Scaled Dot-Product Attention function, mapping a query and a set of key-value pairs to an output, where the query ($Q$), keys ($K$), values ($V$), are learnable representative vectors for the instances in the input sequence with dimensions $d_k$, $d_k$, and $d_v$, respectively. The output of a self-attention module is computed as a weighted average of the values, where the weight assigned to each value is computed by a similarity function of the query and the corresponding key after applying a softmax function~\cite{Vaswani2017}. More specifically, the attention values for a set of queries are computed simultaneously, packed together into a matrix $Q$. The keys and values are similarly represented by matrices $K$ and $V$. The output of the attention Scaled Dot-Product Attention function is computed~as
\vspace{-0.1in}
\begin{equation}
    Attention(Q,K,V) = softmax(\frac{QK^{T}}{\sqrt{d_k}})V
    \label{eq:attention}
    \vspace{-0.08in}
\end{equation}
where $K^{T}$ is the transpose of the matrix $K$.
It is also beneficial to linearly project the queries, keys, and values $h$ times with different learnable
linear projections to vectors with $d_k$, $d_k$ and $d_v$ dimensions, respectively, before applying the attention function. On each of the projected versions of queries, keys, and values, the attention function is performed in parallel, resulting in several $d_{v}$ dimensional output values. These values are then concatenated and once again linearly projected via a fully-connected layer. This process is called ``Multi-Head Attention (MHA)" which helps the model to jointly attend to information from different representation subspaces at different positions~\cite{Vaswani2017}. The output of the MHA module is calculated as
\vspace{-0.05in}
\begin{align}
    &MHA(Q,K,V) = Concat(head_1, \cdots, head_h)W^O, \nonumber \\
    &head_i = Attention(QW_i^Q,KW_i^K,VW_i^V),
    \label{eq:mha}
\end{align}
where the projections are achieved by parameter matrices $W_i^Q \in \mathbb{R}^{d_{model}\times d_k}$, $W_i^K \in \mathbb{R}^{d_{model}\times d_k}$ ,$W_i^V \in \mathbb{R}^{d_{model}\times d_v}$, and $W^O \in \mathbb{R}^{{hd_v}\times d_{model}}$.
%============================================================
\begin{figure}[t!]
    \centering
    \includegraphics[scale=.4]{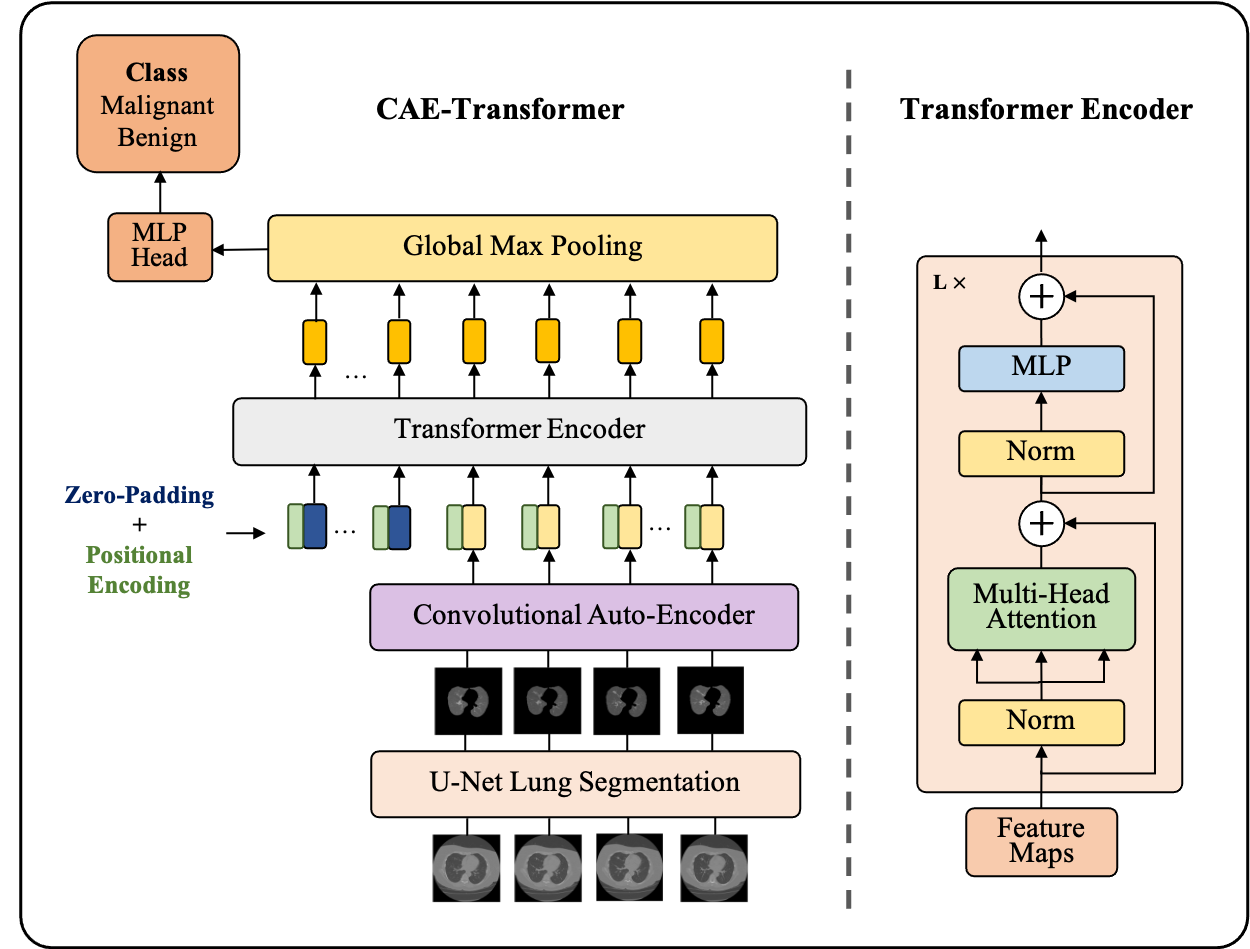}
    \caption{Left: Pipeline of the CAE-Transformer, Right: Architecture of the Transformer Encoder}
    \label{fig:cae-transformer}
    \vspace{-0.2in}
\end{figure}
%============================================================
%============================================================
\begin{table}[t!]
    \centering
     \caption{Results obtained by the CAE-Transformer and its counterparts.}
     \vspace{0.05in}
    \begin{tabular}{|c|c|c|c|c|}
    \hline
        \textbf{Model} & \textbf{Accuracy} & \textbf{Sensitivity} & \textbf{Specificity} & \textbf{AUC}  \\
    \hline
        Reference~\cite{Oikonomou2019} & 81.00\% & 80.00\% & 81.80\% & 0.91 \\
    \hline
        GMP-FC & 84.02\% & 87.00\% & 80.67\% & 0.90 \\
    \hline
        GAP-FC & 83.18\% & 85.33\% & 80.67\% & 0.90 \\
    \hline
        CAE-LSTM & 84.92\% & 85.00\% & 84.33\% & 0.84 \\
     \hline
        \makecell{3D-CNN} & 84.10\% & 85.33\% & 82.67\% & 0.89 \\
     \hline
        \makecell{ResNet18} & 84.55\% & 84.67\% & 84.33\% & 0.91 \\
    \hline
        \makecell{ResNet34} & 85.76\% & 86.67\% & 84.34\% & \textbf{0.94} \\
    \hline
        \makecell{SE-ResNet18} & 84.92\% & \textbf{88.67\%} & 81.00\% & 0.93\\
   \hline
        \makecell{SE-ResNet34} & 84.10\% & 83.00\% & 84.67\% & 0.87 \\
    \hline
        \makecell{CAE-Transformer\\(Concatenation)} & 85.83\% & 83.00\% &\textbf{88.33\%} & 0.92 \\
     \hline
        \makecell{CAE-Transformer\\(GAP)} & 85.83\% &  87.00\% & 84.67\% & 0.88 \\
    \hline
        \makecell{CAE-Transformer\\(GMP)} & \textbf{87.73\%} & \textbf{88.67\%} & 86.33\% & 0.91\\
    \hline
    \end{tabular}
    \vspace{-0.2in}
    \label{tab:results}
\end{table}
%%%%%%%%%%%%%%%%%%%%%%%%%%%%%%%%%%%%%%%%%%%%%%
%============================================================
\subsection{CAE-Transformer}
%============================================================
Fig.~\ref{fig:cae-transformer} illustrates the pipeline of the CAE-Transformer framework, along with the architecture of a transformer encoder in which LN represents the layer normalization and MLP stands for Multi-Layer Perceptron. The transformer model used in the CAE-Transformer framework is adopted from the transformer encoder proposed in References ~\cite{Vaswani2017,Dosovitskiy2020} and modified for the task at hand.
The CAE-Transformer is constructed by stacking $3$ transformer encoder blocks on top of each other with the projection dimension of $256$,  key and query dimensions of $128$, and $5$ number of heads in each MHA module. Finally, the features obtained by the stack of transformer encoders from all input instances (slices) are passed to a GMP layer, followed by two Fully-Connected (FC) layers with $32$ and $2$ neurons, respectively, to provide the final predictions. The last fully-connected layer uses a softmax activation function to produce probability scores. Dropout layers are also incorporated to prevent the model from getting over-fitted. In addition, following the literature~\cite{Vaswani2017,Dosovitskiy2020},  a Positional Embedding (PE) layer is incorporated into the model to add information about the position of instances in the input sequence  to the CAE-generated features. It is worth noting that as the number of slices with the evidence of a nodule varies for different subjects (from $2$ to $25$ slices per nodule), we have taken the maximum number of slices in our dataset (i.e., $25$ slices) and zero-padded the input sequences based on this number, so that all of them have the same dimension of ($25,256$).
%Furthermore, unlike existing transformers, we have not used a global pooling layer to aggregate the feature maps obtained by the last transformer encoder. Instead, we concatenated all the sequential features generated by the last transformer encoder and fed the result to the subsequent FC model to provide the final outcome.

%The following equations describe how the CAE-Transformer's output is obtained as
%
%\begin{align}
 %   &({s_1}', {s_2}', \cdots,  {s_{c_i}}') = \text{\textit{U-Net}}(s_1, s_2, \cdots, s_{c_i}), &\quad i=1\cdots N \nonumber \\
  %  &(f_1, f_2, \cdots,  f_{c_i}) = CAE(({s_1}', {s_2}', \cdots,  {s_{c_i}}'), &\quad i=1\cdots N \nonumber \\
   % &z_0 = ZeroPad(f_1, f_2, \cdots,  f_{c_i}),  + PE &\quad i=1\cdots N \nonumber \\
   % &{z_l}' = MSA(LN(z_{l-1})) + z_{l-1}, &\quad l=1\cdots L \nonumber \\
   % &z_l = MLP(LN({z_l}')) + {z_l}', &\quad l=1\cdots L \nonumber \\
   % &o = LN(z_L) ,\nonumber  \\
   % &x = GMP(o_1, o_2, \cdots, o_{25}), \nonumber \\
   % &y = MLP(x),
   % \label{eq:cae_transformer}
%\end{align}
%
%where $s$ denotes the original CT slices, ${s}'$ represents the segmented CT images, $c_i$ signifies the number of slices with the evidence of a nodule in the case $i$, $f$ represents the CAE-generated feature maps corresponding to the CT images, $MSA$ denotes the Multi-Head Self-Attention module, and $l$ shows the $l_{th}$ MSA layer. The number $25$ indicates the maximum number of slices with the evidence of a nodule per subject in this study, and $y$ is the final prediction.

%OOOOOOOOOOOOOOOOOOOOOOOOOOOOOOOOOOOOOOOOOOOOOOOOOOOOOOOOOOOOOO
\vspace{-0.1in}
\section{Results} \label{sec:pagestyle}
%OOOOOOOOOOOOOOOOOOOOOOOOOOOOOOOOOOOOOOOOOOOOOOOOOOOOOOOOOOOOOO
We evaluated the performance of the proposed CAE-Transformer framework using the 10-fold cross-validation method. The CAE model is pre-trained using a batch size of $128$, learning rate of $1e-4$ and $200$ epochs. The best model on the randomly sampled $20\%$ of the dataset was selected as the best model. The model was then fine-tuned on the in-house dataset using a lower learning rate of $1e-6$ and $50$ epochs. To fine tune the final CAE, only the middle fully-connected layer and its previous and next convolution layers were trained, while the other layers were kept unchanged.
The CAE-generated features were then used to train the transformer encoder. The transformer was trained using a learning rate of $1e-4$, batch size of $64$, and $200$ epochs. We also employed label smoothing with the $\alpha=0.05$~\cite{Szegedy_2016_CVPR}.
The results obtained by the CAE-Transformer are presented in Table~\ref{tab:results}.
We have compared the performance of the CAE-Transformer with the results obtained by the original model proposed in Reference~\cite{Oikonomou2019}.
We have further compared the CAE-Transformer with non-transformer-based alternative models, referred to as GMP-FC and GAP-FC,  by aggregating the CAE-generated feature maps using GMP and GAP, respectively,  followed by a stack of fully connected and batch normalization layers. The best experimental results for such models were obtained by utilizing $4$ fully connected layers with $128$, $128$, $32$, and $2$ neurons, respectively.

We have also compared performance of the CAE-Transformer with its deep learning-based counterparts. First,  the CAE-LSTM is developed by replacing the transformer blocks with a stack of LSTM layers, while keeping the hyper-parameters and complexity the same.
Then,  we trained a custom 3D-CNN model, containing $4$ convolution, $4$ max-pooling, $1$ batch normalization, and $1$ dropout layers, followed by $2$ FC layers.  We also modified the last layers of the ResNet18~\cite{He2015}, ResNet34~\cite{He2015}, SE-Resnet18~\cite{Hu2017}, and SE-Resnet34~\cite{Hu2017} to be compatible with the classification task at hand and trained them on the same dataset.  Such 3D CNN-based models are the building blocks of many frameworks developed in the field of medical image processing~\cite{Gu2021,Mohammadi2021}.  It is worth noting that larger and deeper networks did not perform well on the small dataset used in this study. As the last experiment,  we investigated the effects of different pooling layers on top of the network.  More specifically,  we replaced the GMP layer by a GAP and concatenation function in separate experiments. It is worth mentioning that other models proposed in the literature are developed based on annotated tumor patches from different datasets, which makes re-training those models on our dataset impossible.  As such, comparison with those studies is not included.
The experimental results provided in Table~\ref{tab:results} demonstrate that deep learning-based models outperform the original radiomics and machine learning-based model, while the proposed CAE-Transformer achieves the best performance among the developed frameworks.
%The results also demonstrate that incorporating the GAP, GMP, and PE into the model deteriorates the performance in our study. In the case of PE, we suspect that understanding additional positional relations can be challenging when the model is trained on a relatively small dataset like ours. The recently published study in~\cite{Valanarasu2021} also reports the same issue when a small training dataset is used to train a transformer. It is worthy of note that increasing the complexity of the model could improve the performance when GAP, GMP, and PE were included. The improvements, however, were minor and could not reach the CAE-Transformer's results.

%Experimental results showed that deep learning-based models improve the results achieved by the study performed in~\cite{Oikonomou2019} based on the histogram-based and radiomics features, while the CNE-Transformer provided the highest improvement. More specifically, the CNE-Transformer improved the accuracy from $81.0\%$ to $87.73\%$, sensitivity from $80.0\%$ to $88.67\%$, and specificity from $81.8\%$ to $86.33\%$, while achieving the same high AUC value of $0.91$.

%OOOOOOOOOOOOOOOOOOOOOOOOOOOOOOOOOOOOOOOOOOOOOOOOOOOOOOOOOOOOOO
\section{Conclusion}
%OOOOOOOOOOOOOOOOOOOOOOOOOOOOOOOOOOOOOOOOOOOOOOOOOOOOOOOOOOOOOO
In conclusion, we have proposed an automated transformer-based framework, referred to as the ``CAE-Transformer", to enhance the performance of existing models aiming to predict the invasiveness of lung adenocarcinoma subsolid nodules from 3D CT scans regardless of technical acquisition settings.
%The experimental results on our in-house dataset showed significantly improvements made by the proposed CAE-Transformer compared to the original radiomics and machine learning-based model~\cite{Oikonomou2019}.
%by increasing the accuracy by $6.73\%$, sensitivity by $8.67\%$, and specificity by $4.53\%$. The CAE-Transformer is also capable of capturing global inter-slice relations in a volumetric CT scan while requiring less computational resources compared to RNN and LSTM.
%We have also investigated the effects of GMP,  GAP, and PE in our model and realized that such components of a conventional transformer are not beneficial for the task at hand, especially in our case where the training dataset is relatively small.
We would like to mention that achieving a clinically applicable deep learning-based solution requires more experiments and research studies in this field, and we believe that the proposed framework is one step forward towards achieving such clinically applicable frameworks. As shown in the comparison results, the proposed CAE-Transformer performs far better than the original models developed in Reference~\cite{Oikonomou2019}, which is a study on the same dataset based on machine learning approaches and radiomics features extracted from manually annotated tumors. The experimental results of this study further encourage researchers in the medical signal/image processing society to adopt more deep learning-based models, in particular, transformer-based models to target similar analytic and predictive tasks. In future works, we will be collaborating with our partners in medical centers to increase the size and diversity of the dataset and target the three-way SSN classification task.  Furthermore, we would like to investigate the effects of incorporating radiomics and morphological features into the CAE-Transformer.

\label{sec:typestyle}
%\vfill
%\pagebreak

%\vfill
%\pagebreak
\vspace{-0.05in}
\bibliographystyle{IEEEbib}
\bibliography{refs}

\end{document}